\documentclass[aip,apl,amsmath,amssymb,reprint,superscriptaddress]{revtex4-1}
\usepackage{graphicx}
\usepackage{amsmath}
\usepackage{dcolumn}
\usepackage{bm}
\usepackage{bbold}
\usepackage{physics}
\usepackage{scalerel}

\bibstyle{apsrev4-1}

\begin{document} 
    
\title{Rigorous ESR spectroscopy of $Fe^{3+}$ impurity ion with oxygen vacancy in ferroelectric $SrTiO_3$ crystal at 20 mK}

\author{M. A. Hosain}
\email{akhter361@yahoo.co.uk}
\affiliation{ARC Centre of Excellence for Engineered Quantum Systems, School of Physics, University of Western Australia, 35 Stirling Highway, Crawley WA 6009, Australia.}

\author{J-M. Le Floch}
\affiliation{MOE Key Laboratory of Fundamental Physical Quantities Measurement, School of Physics, Huazhong University of Science and Technology, Wuhan 430074, Hubei, China.}
\affiliation{ARC Centre of Excellence for Engineered Quantum Systems, School of Physics, University of Western Australia, 35 Stirling Highway, Crawley WA 6009, Australia.}

\author{J. Krupka}
\affiliation{Department of Electronics and Information Technology, Institute of Microelectronics and Optoelectronics, Warsaw University of Technology, Koszykowa 75, 00-662 Warszawa, Poland.}

\author{M. E. Tobar}
%\email{michael.tobar@uwa.edu.au}
\affiliation{ARC Centre of Excellence for Engineered Quantum Systems, School of Physics, University of Western Australia, 35 Stirling Highway, Crawley WA 6009, Australia.}

%\date{\today}

%%%%%%%%%%%%%%%%%%%%%%%%%%%%%%%%%%%%%%%%%%%%%%%%%%%%

\begin{abstract}
Impurity $Fe^{3+}$ ion  electron spin resonance (ESR) spectroscopy using multiple dielectric modes in a $SrTiO_3$ dielectric resonator has been performed with a tuneable DC magnetic field of up to $1.6~T$. The $Ti^{4+}(d^0)$ ion is substituted by $Fe^{3+}$ ion forming $FeO_6$ octahedral complex with an iron-oxygen-vacancy $(Fe^{3+}-V_O)$. In such a metal-ligand complex, a giant g-factor of $g_{\scriptscriptstyle\parallel F} = 5.51$ was observed in the ferroelectric phase at $20~mK$. The the change of $Fe^{3+}$ ion center-symmetry in the $FeO_6$ complex as a soft-mode characteristics of ferroelectric phase transition and the influences of iron-oxygen-vacancy $(Fe^{3+}-V_O)$, interactively sensitive to asymmetry in the octahedral rotational parameter $\Phi$ in $SrTi0_3$. 
\end{abstract}

\maketitle

\subsection{Introduction:}

 ESR spectroscopy, which implements dielectric crystal resonators operating with multiple modes has been proven as an effective method for study on impurity paramagnetic ions' unpaired electron spin states$\cite{Farr,Karim,Buluta}$. This process assimilate a quantum-hybrid system by coupling unpaired electron spins with photonic modes of the dielectric crystal resonator$\cite{KurizkiHybSy,RoadMapHybSy,Bensky2011,MagnVibrSpinCrossBO}$. The multi-mode ESR spectrum works as a direct probe providing information of electronic states of paramagnetic impurity ions. In this work, the $Fe^{3+}(d^5)$ ion has been detected in the site of the $Ti^{4+}(d^0)$ ion of the dielectric single crystal $SrTiO_3(STO)$ at $20~mK$. The properties of $3d^n$-metal-ligand complex in crystals are closely related to the central $3d^n$ ion and the ligand ion revealing interesting physical characteristics. The variation of position along the central $Fe^{3+}$ ion of $3d^5~FeO_6$ complex have received an increasing amount of attention. The effect of ferroelectric (FE) phase transition on ESR due to  such kind of central ion displacement reveals important characteristics of unpaired electron quantum states. At the temperature $20~mK$, the crystal is of rhombic symmetry in the ferroelectric (FE) phase deforming from cubic octahedral structure. This structural anisotropy plays a vital role in the mechanism of ESR. Optical spectroscopy and X-ray diffraction (XRD) results are available providing localization information of the $Ti^{4+}(d^0)$ ion in this type of crystal$\cite{DisplaceTi,PrepSTO}$. In this experimental study of ESR, we justify the multi-valance $Fe$ ion's measured spin-Hamiltonian parameters along with site symmetry considering delocalization and structural anisotropy.

Field confinement in $SrTiO_3$ is very high due to it extremely high permittivity at low temperatures, this allows only the loss mechanisms of the crystal itself to determine the resonator Q-factor$\cite{Krupka,LeFloch,WG,Krupka1}$. Required high Q-factors can be maintained due to no significant metal cavity or radiation losses, necessary to detect paramagnetic impurity ions of concentration level parts per million (ppm) to parts per billion (ppb) at $20~mK$. Low concentration of ions requires high sensitivity providing accurate information about the nature of the paramagnetic species, and its immediate surroundings which affect on the anisotropy, fine and hyperfine splitting.
 
Single crystal $SrTiO_3$ octahedral metal-ligand complex has a high crystal field (CF),  with triplet($t_{2g}$)-doublet($e_g$) orbital splitting of transition metals of about $10D_q=20200~cm^{-1}$ $\cite{STO-CrystField,STO-CrystField1}$. $SrTiO_3$ crystallizes in a perovskite structure, space group Pm3m with a lattice constant of $3.905$~\r{A} in cubic symmetry, and the $Ti^{4+}-O^{2-}$ distance is $1.952$~\r{A}\cite{STO-CrystField1}. Rowley et al suggested that the FE phase transition temperature of $SrTiO_3$ was higher than $50K$ confirming its phase transition from paraelectric (PE) to FE$\cite{QCriticality}$. In FE phase transitions, the average off-center displacement of the paramagnetic metal ion is about $0.1$~\r{A} compared to lattice constant of the order $4$~\r{A}$\cite{MagnJT-Green}$. To cause a $100~cm^{-1}$ energy change on average a $0.01$~\r{A} off-center displacement$\cite{CuHole}$ would be required.  In this crystal field symmetry with little distortion of octahedral structure as tetragonal symmetry or ultimately rhombohedral symmetry; the spin-Hamiltonian parameters of paramagnetic ions were measured with respect to the basis of a crystal field model at low magnetic field ($\leq 15~mT$), and at high magnetic field by implementing P-band to X-band multiple mode frequencies.

\begin{figure}[t!]
\centering
\includegraphics[width=3.5in]{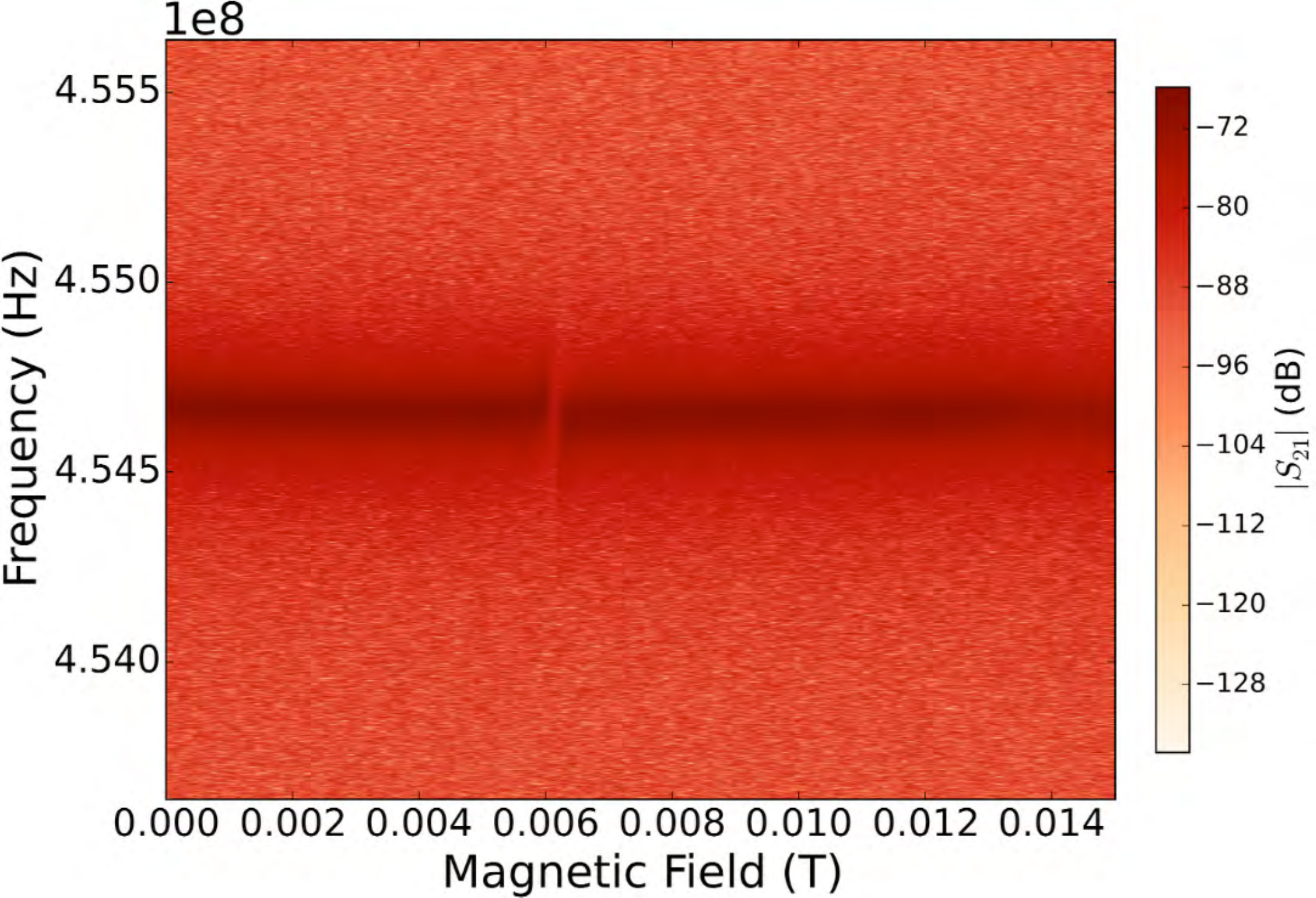}
\caption[ Density plot of spectrum at mode frequency 0.4546 GHz]{\label{plot45} Density color plot with $ESR$ spectrum transmission $S_{21}$ $Fe^{3+}$ at mode frequency $0.4546~GHz$ in $SrTiO_3(STO)$ crystal at $20~mK$ temperature.}
\end{figure}
\begin{figure}[t!]
\includegraphics[width=3.5in]{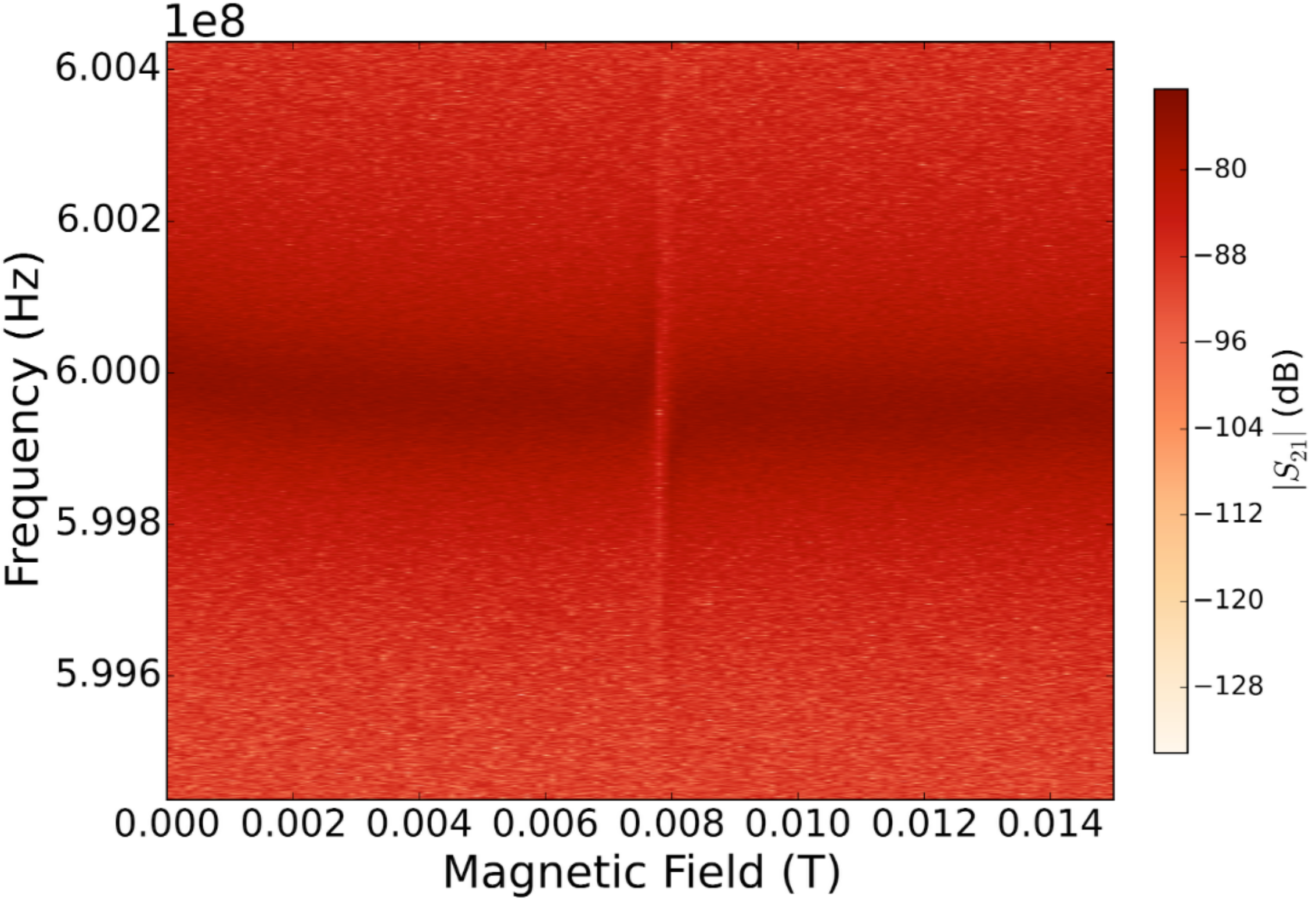}
\caption[ Density plot of spectrum at mode frequency 0.5993 GHz]{\label{plot59} Density color plot with $ESR$ spectrum transmission $S_{21}$ $Fe^{3+}$ at mode frequency $0.5993~GHz$ in $SrTiO_3(STO)$ crystal at $20~mK$ temperature.}
\end{figure}
\begin{figure}[t!]
\includegraphics[width=3.5in]{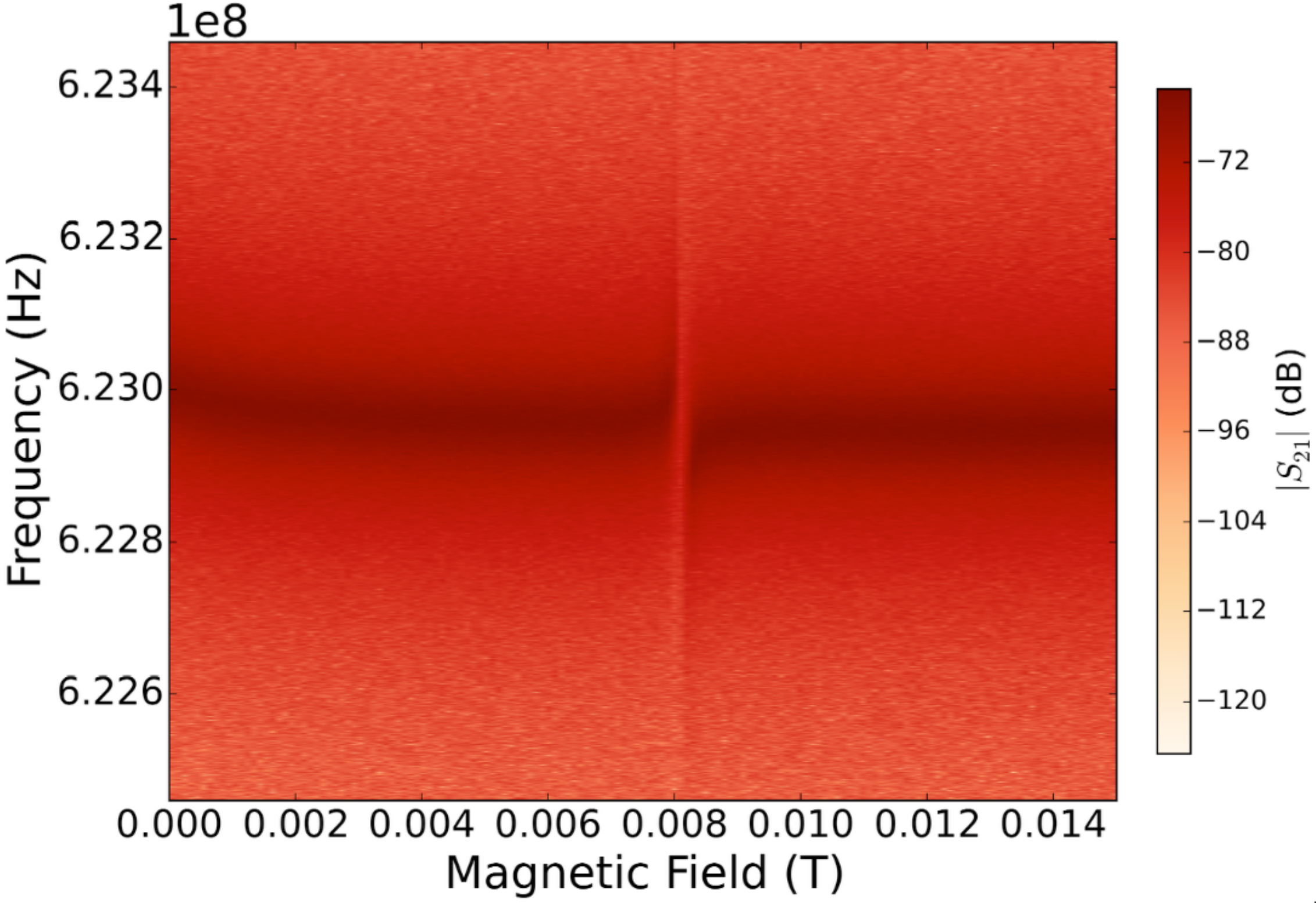}
\caption[ Density plot of spectrum at mode frequency 0.6228 GHz]{\label{plot62} Density color plot with $ESR$ spectrum transmission $S_{21}$ $Fe^{3+}$ at mode frequency $0.6228~GHz$ in $SrTiO_3(STO)$ crystal at $20~mK$ temperature.}
\end{figure}

\subsection{Electron Spin Resonance (ESR) spectroscopy:}

For paramagnetic metal ions' unpaired electrons in the condensed phase, the CF splitting and the spin-orbit coupling (SOC) are responsible for electronic interactions showing effective spin. In case of strong CF of $SrTiO_3$, the combined effect of these two interactions may remove the orbital degeneracy of the energy levels of most metal ions completely, leaving a non-degenerate ground state with quenched orbital momentum. In the symmetry less than cubic, fine structure term $D\{S_z^2 - \frac{1}{3}S(S+1)\}$ will generally be present.

The $3d^6$ and $3d^5$ electron configuration  of $Fe^{2+}$ ion and $Fe^{3+}$ ion orbital splitting in octahedral crystal field have the ground state as an orbital triplet ${^2}T_{2g}$. Low-spin in high crystal field follows the treatment as given by McGravey $\cite{McG,McGarvey1}$. Hence, the $Fe^{2+}$ ion in this symmetry of $SrTiO_3$ high CF has effective spin $S=0$ forming a spin singlet state of ${^2}T_{2g}$ orbitals. Under these conditions, only a small temperature independent susceptibility is observed for the ground state$\cite{AbragamESR}$. In the case of $Fe^{3+}$ ion low-spin state, the weak field sextet state $^{6}S_\frac{5}{2}$ is altered to spin state of effective spin $S=\frac{1}{2}$ attaining a lower energy in the lower symmetry of crystal. Within the triplet orbital ground state of fictitious orbital $l=1$, we have only one electron spin-orbit coupling $\zeta(\bold{l.S})$. For $S=\frac{1}{2}$, fine structure term becomes zero. Hence, the spin-Hamiltonian for effective spin $\bold{S}$ in applied external magnetic field $\bold{B}$ along z-axis may be described as$\cite{AbragamESR}$
 \begin{eqnarray}
\label{eq:HamilSTO}
H= g_{\scriptscriptstyle\parallel}\beta B S_z +E(S_x^2 + S_y^2) + A_{\scriptscriptstyle\parallel}S_zI_z
\end{eqnarray} 
 
where $g_{\scriptscriptstyle\parallel}$ is the electron spin parallel g-factor, $E$ is the energy term of rhombic distortion and $I$ is the nuclear spin, $A$ is the hyperfine constant.\

In case of low magnetic field $\leq 15~mT$, the $ESR$ spectroscopy is challenging in the state of zeeman transition energy less than hyperfine coupling energy. The spins $\bold{S}$ and $\bold{I}$ are strongly coupled, and, therefore, the spin system can not be described in the basis $\ket{M_S,M_I}$ with electron magnetic moment quantum number $M_S$ and nuclear magnetic moment quantum number $M_I$. The basis $\ket{F,M_F}$ has to be used, where $F=S+I$ is the total quantum number and $M_F$ the magnetic moment quantum number of total spin with g-factor as $g_F$ in such a state$\cite{AbragamESR}$. Hence, the description of the wave functions of the system for $I=\frac{1}{2}$ ion in $SrTiO_3$ in this basis are$\cite{STO-FeBasis,STO-EsrEffectLowField}$-   
\begin{eqnarray}
\label{eq:HamilSTObasis}
\lefteqn{\ket{1,1}_F= \ket{\frac{1}{2},\frac{1}{2}}_{SI}} \nonumber\\&&
\ket{1,0}_F= cos~\varphi\ket{\frac{1}{2},-\frac{1}{2}}_{SI} + sin~\varphi \ket{-\frac{1}{2},\frac{1}{2}}_{SI}\nonumber\\&&
\ket{1,-1}_F= \ket{-\frac{1}{2},-\frac{1}{2}}_{SI}\nonumber\\&&
\ket{0,0}_F= -sin~\varphi \ket{\frac{1}{2},-\frac{1}{2}}_{SI} + cos~\varphi \ket{-\frac{1}{2},\frac{1}{2}}_{SI}
\end{eqnarray}
where the angle $\varphi = tan^{-1}\bigg\{\frac{(\gamma_{\scaleto{S}{2.5pt}} - \gamma_{\scaleto{I}{2.5pt}})B + \sqrt{A^2-(\gamma_{\scaleto{S}{2.5pt}} - \gamma_{\scaleto{I}{2.5pt}})^2B^2}}{A} \bigg\}$, also $\gamma_{\scaleto{S}{3pt}}$ and $\gamma_{\scaleto{I}{3pt}}$ is the gyromagnetic ratio of electron and proton respectively.\\

For $ESR$ spectroscopy, the cylindrical $STO$ crystal specimen of diameter $3.27~mm$ and height $3.66~mm$ was used. The modes utilised were in general low order with azimuthal variation $m=0$ to $3$ determined by computer simulation software based on the Method of Lines (MoL)$\cite{LeFloch}$. The crystal was inserted centrally in an oxygen-free cylindrical copper cavity, and the modes were excited as a dielectric resonator with a Vector Network Analyzer $(VNA)$ sensing the modes in transmission by measuring $S_{21}$. By matching measured and simulated mode frequencies, the relative permittivity of the crystal was estimated to be $316\pm 0.5$ at room temperature $295\pm5$K.

The crystal loaded cavity was cooled in a dilution refrigerator $(DR)$ to less than $20~mK$. At this temperature, all the excited mode frequencies were confined within the range of frequency $0.4 - 1.2~GHz$ due to the increase of relative permittivity to the order of $10^4$. High sensitivity of $ESR$ spectroscopy is required for such a very low concentration of impurity ions in this nearly pure $SrTiO_3$ crystal. 
Different process are devoted in measuring sensitivity of different type of resonator of wide range of frequency with varieties of probing system$\cite{AnninoQL,LongoQL,ColligianiQL,AndersMK,YapMK,SLA}$. Benmessai et al.$\cite{Karim}$ described a concentration level measurement process of $Fe^{3+}$ impurity ion exciting modes at millikelvin temperatures in sapphire. Anders et al$\cite{AndersMK}$ described a single-chip electron spin resonance detector operating at $27~GHz$.\

 Practically, microwave-power and other terms are kept constant, the required minimum number of impurity ion follows the proportionality$\cite{CharlesESR}$ $N_{min} \propto \frac{1}{\omega Q_L}$ for detection of ESR transition spectrum, and is estimated generally as:
 \begin{eqnarray}
\label{eq:Nmin}
 N_{min}=\Big(\frac{3 K_B V_sT_s}{g^2\beta^2\mu_\circ S(S+1)}\Big)\Big(\frac{\Delta \omega}{\omega}\Big)\Big(\frac{1}{\eta Q_L}\Big)\Big(\frac{P_n}{P}\Big)^\frac{1}{2}~~~~~~~   
\end{eqnarray} 
 Where $V_s$ is the mode volume, $T_s$ is the sample temperature, S is the electron effective spin, $g_e$ is the electron g-factor, $\beta$ is the Bohr electron magneton, $\mu_\circ$ is the magnetic permeability of free space, $\omega$ is the resonance frequency, $\eta$ is the filling factor, $P_n$ is the noise power, $P$ is the microwave input power, and $\Delta \omega$ is the width of aggregated spin frequency at resonance which is depended on the shape-function $f(\omega)$ normalized as $\int_0^\infty f(\omega)\partial\omega=1$ for a wide range of Larmor precession $(\omega_L)$ of magnetic dipoles$\cite{AbragamESR}$. Significant output (transmission) occurs only at resonance in a very narrow frequency width $\Delta \omega$ in the region $\omega\approx\omega_L$ at ESR$\cite{SLA,SLACu}$. This frequency width is about $300000~Hz$ in the ferroelectric anisotropy of the crystal $SrTiO_3$ crystal (see Figs.\ref{plot45}, \ref{plot59} and \ref{plot62}), and $Q$-factor of the selected modes were about $1000$ at $20~mK$ temperature. This frequency width is quite large due to ferroelectric anisotropy effect. For $Fe^{3+}$ ion with effective spin $S=\frac{1}{2}$ in this $SrTiO_3$ specimen, calculation using $Eq.\ref{eq:Nmin}$, the concentration level of ion about 1 ppm is required at $20~mK$ in a state $\frac{P_n}{P}=1$.\
 
Ten modes with a frequency range of $0.4~GHz$ to $1.2~GHz$, and thus monitor the transition spectrum by considering mode magnetic field vector at an angle of $\theta$ to the direction of the applied $DC$ magnetic field $B$. Such a mode field vector interaction in ESR is represented by the quantum-mechanical operator\cite{STO-FeBasis,STO-EsrEffectLowField} $(cos~\theta S_z +\frac{1}{2} sin~\theta S_\pm)$ composed with ladder operator $S_\pm$.\

The applied $DC$ magnetic field $\bold{B}$ between $0~T$ to $1~T$ was varied through the use of computer control, with a step width of $4\times 10 ^{-4}~T$. Each mode is scanned for a period of five seconds at each step of magnetic field. This slow sweep of magnetic field was applied under control of an in-house $MATLAB$ program to avoid heating above $20~mK$, with the microwave input power of $-60~dBm$. To avoid the addition of thermal noise from room temperature, a $10~dB$ microwave attenuator was used at $4~K$ stage and another one at $1~K$ stage of the $DR$. Also, a $20~dB$ attenuator was added at $20~mK$ stage of the $DR$. These cold stage attenuation plus the use of a low noise temperature cryogenic amplifier after the resonator ensures good enough signal to noise ratio $(SNR)$. From this measured multi-mode $ESR$ spectrum characteristics, we were able to identify the types of paramagnetic impurities present in the crystal including spin-Hamiltonian parameters.

\subsection{Results and Discussion}

\begin{figure}[b!]
\includegraphics[width=3.5in]{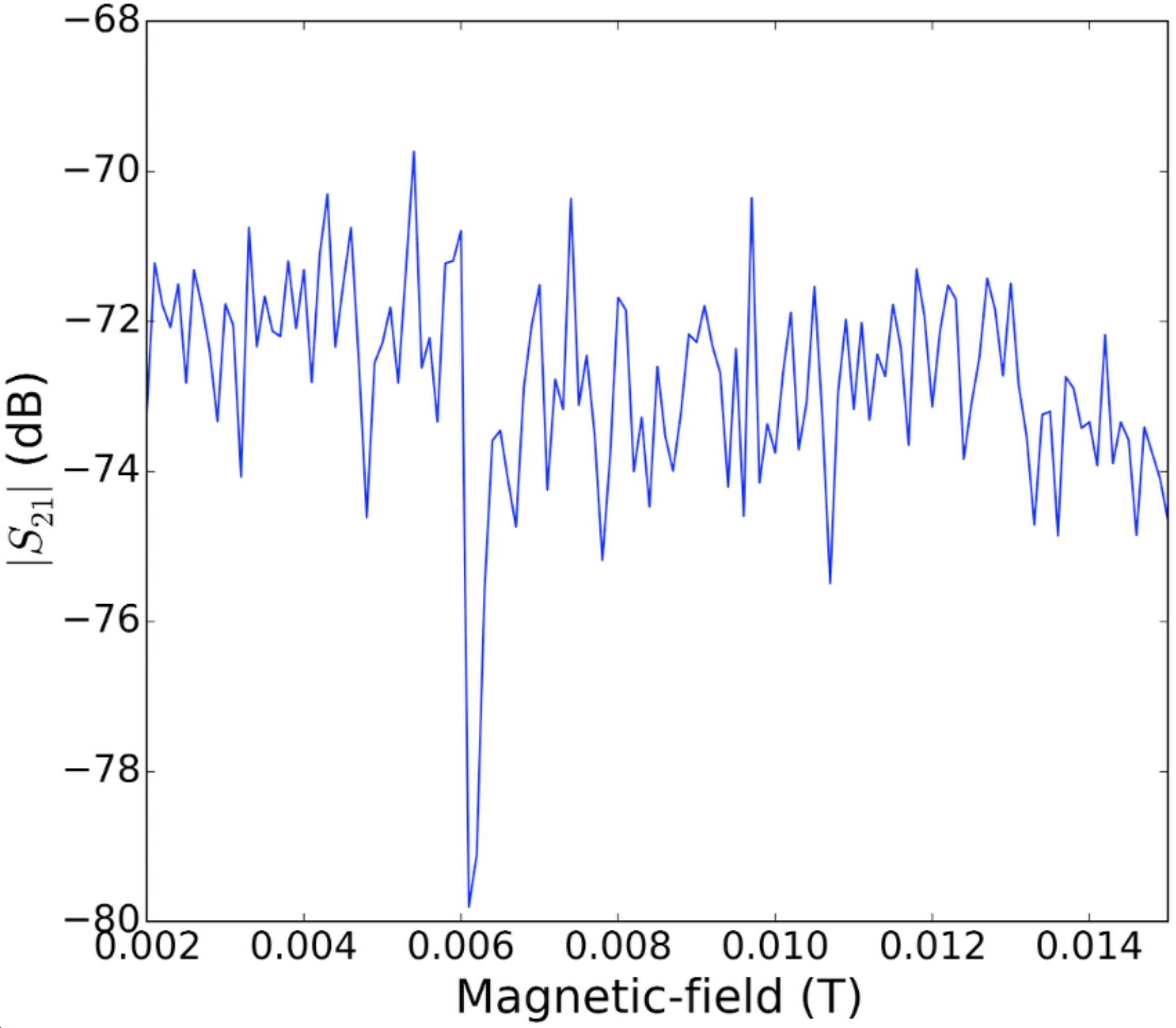}
\caption[Transmission sweep of $ESR$ line of $Fe^{3+}$ at $0.4546~GHz$]{\label{sweep454} Transmission amplitude in dB and line shape of $Fe^{3+}$ ion resonance at $0.4546~GHz$.}
\end{figure}
\begin{figure}[t!]
\includegraphics[width=3.5in]{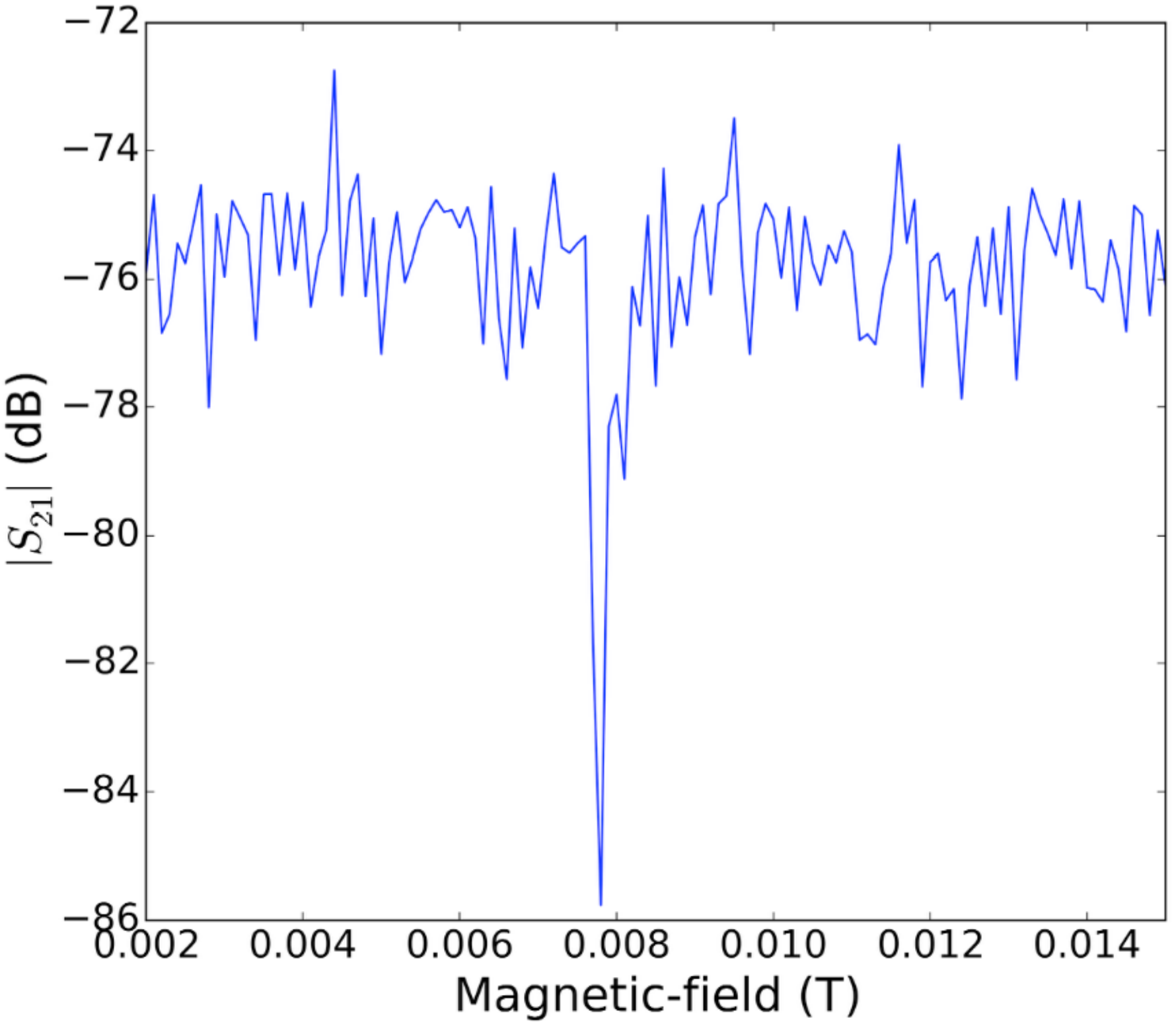}
\caption[ Transmission sweep of $ESR$ line of $Fe^{3+}$ at $0.5993~GHz$]{\label{sweep599} Transmission amplitude in dB and line shape of $Fe^{3+}$ ion resonance at $0.4546~GHz$.}
\end{figure}
\begin{figure}[t!]
\includegraphics[width=3.5in]{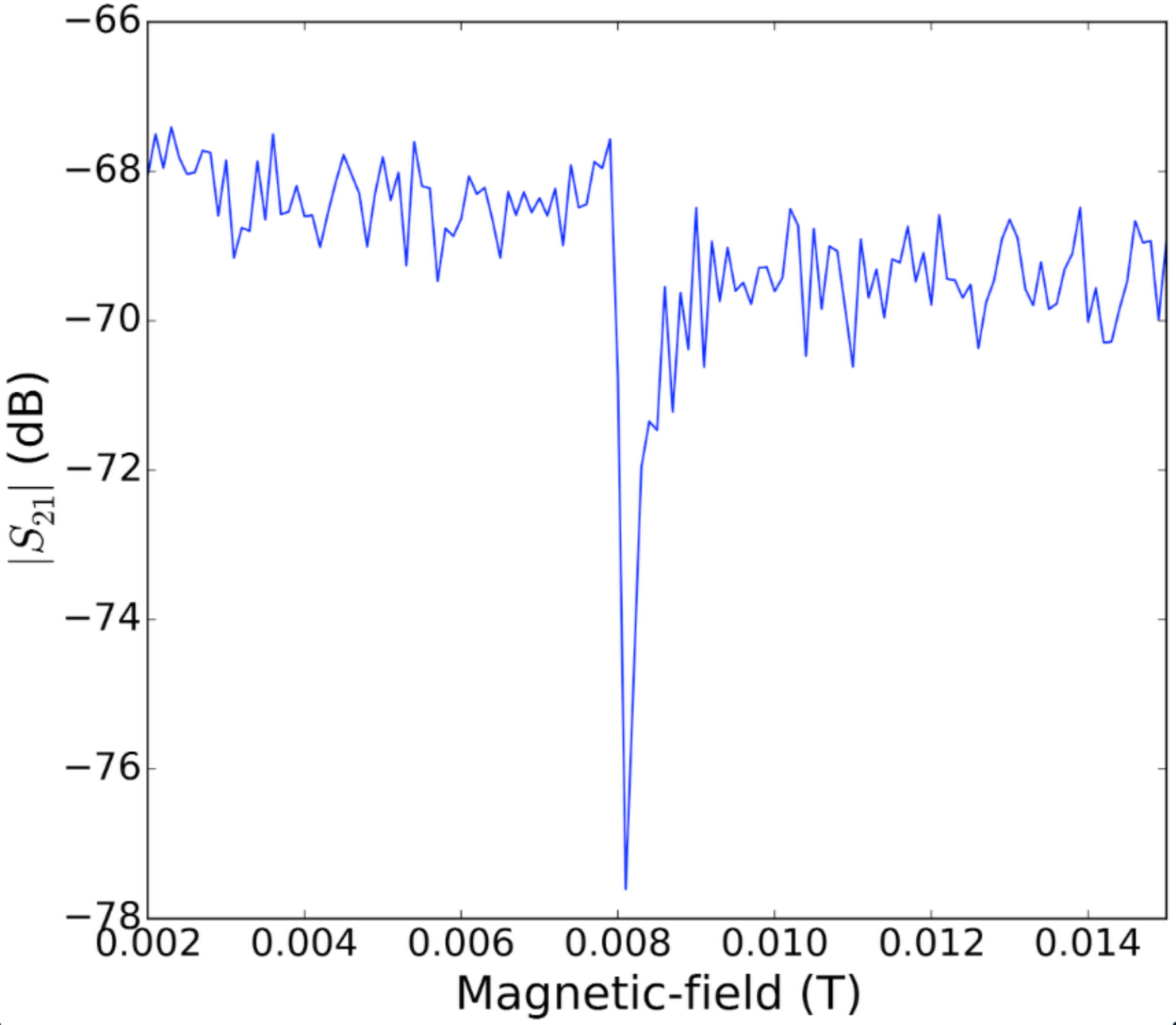}
\caption[ Transmission sweep of $ESR$ line of $Fe^{3+}$ at $0.4546~GHz$]{\label{sweep622} Transmission amplitude in dB and line shape of $Fe^{3+}$ ion resonance at $0.6228~GHz$.}
\end{figure}
\begin{figure}[t!]
\centering
\includegraphics[width=3.5in, height=2.4in]{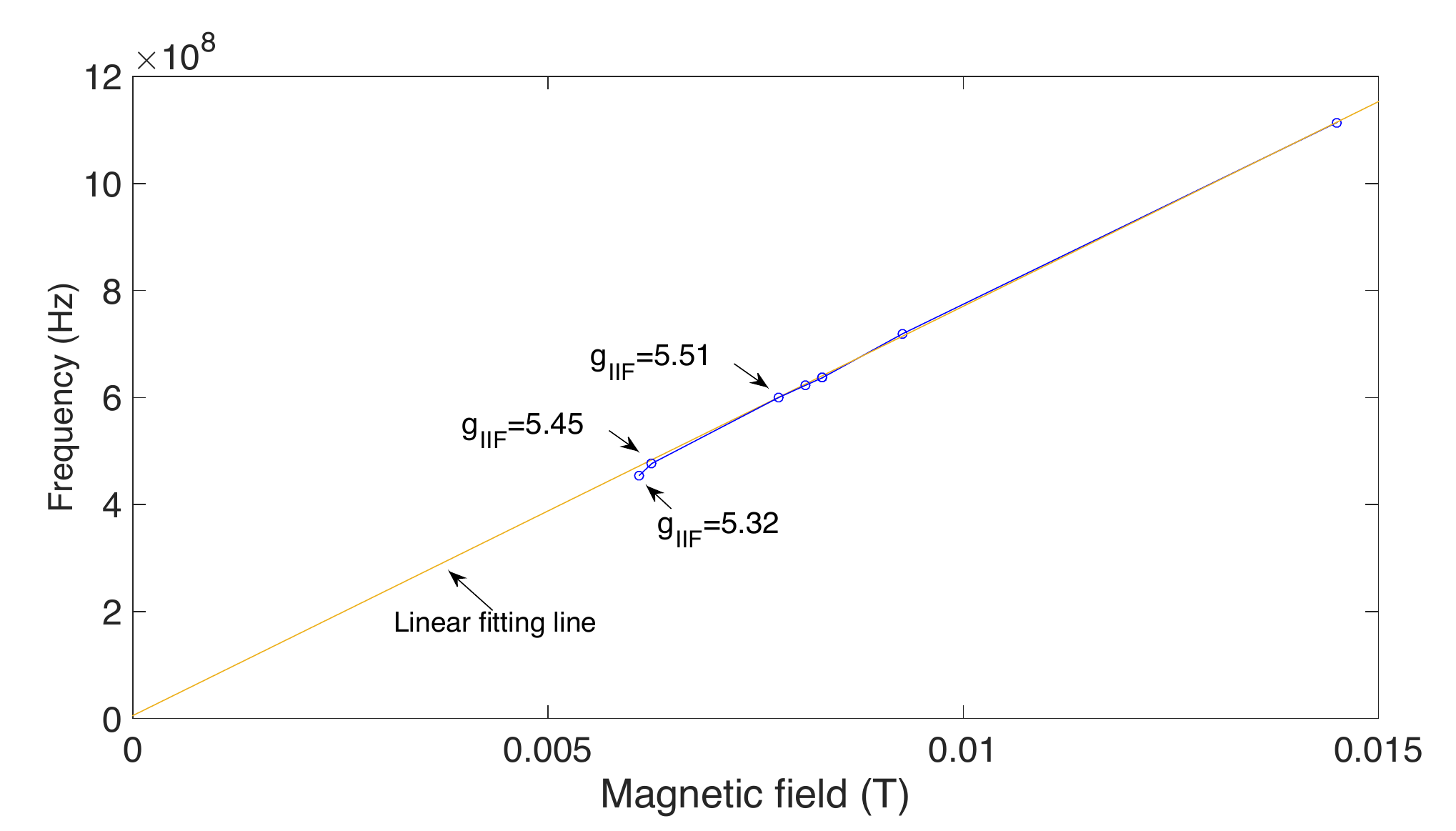}
\caption[ Experimental deduction of effective g-factor of low-spin $Fe^{3+}$]{\label{STOg1}Effective g-factor characteristic graph of low-spin $Fe^{3+}$ ESR line at low-field.}
\end{figure}

The ESR transitions are shown as color density plots in (Fig.\ref{plot45}, \ref{plot59} and \ref{plot62}). All of these transitions are presented in transmission ($S_{21}$) showing the intensity of transition and line shape in (Fig.\ref{sweep454}, \ref{sweep599} and \ref{sweep622}). The calculated effective g-factor $g_{\scriptscriptstyle\parallel F}=5.51$ (see Fig.\ref{STOg1}) indicates a large magnetic moment of electron. The c-axis (z-axis) of the domain was parallel to the $[001]$ crystal direction. For the $Fe^{3+}$ ion in the site of $Ti^{4+}$ ion, an iron-oxygen-vacancy $(Fe^{3+}-V_O)$ is considered$\cite{STO-EsrFE-Muller1,STO-FeVoEsr}$. An angular anisotropy of the $(Fe^{3+}-V_O)$ makes $ESR$ spectrum very sensitive to local rotations $\Phi_l$ of the oxygen octahedron.  M$\ddot{u}$ller and Berlinger$\cite{STO-EsrCriAsyMuller}$ reported that a critical asymmetry in local fluctuations of the rotational parameter $\Phi$ in $SrTi0_3$ for temperature $T<T_c$(FE transition temperature), and $\langle\Phi\rangle\ne 0$ is observed directly by the asymmetry $a_s$ of the paramagnetic resonance lines of $(Fe^{3+}-V_O)$ pair centers. It is postulated that such an asymmetry is a general property of second order $FE$ phase transitions. The results suggest that the $Fe^{3+}$ ions in the $(Fe^{3+}-V_O)$ defect centers are displaced away from the oxygen vacancy $V_O$ by about $0.35$~\r{A}$\cite{STO-FeVoEPR1}$.\

\begin{figure}[b!]
\includegraphics[width=2.1in]{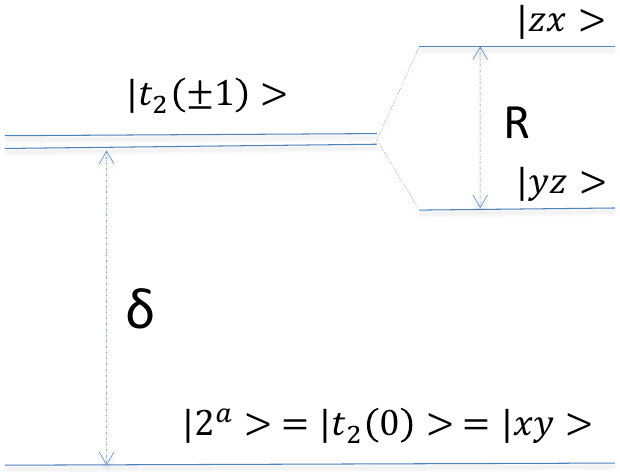}
\caption[ Orbital triplet splitting of $Fe^{3+}$ ion in $SrTiO_3$]{\label{STO-t2g} Orbital splitting of $Fe^{3+}$ ion in tetragonal and rhombic distortion of $FeO_6$ octahedron.}
\end{figure}

Now ignoring the iron-oxygen-vacancy, only considering the FE phase transition geometric deformation of the cubic crystal structure of $SrTiO_3$ to tetragonal structure and ultimately to rhombohedron, the crystal anisotropy in the FE phase also creates nondegeneracy in $T_{2g}$ orbital triplet. All the five electrons of $Fe^{3+}$ ion at the $Ti^{4+}$ site are in ${^2}T_{2g}$ orbitals, and the configuration of one electron approximation is ascribed to a single hole in a filled $d$-shell of orbital triplet. The basis states in the triplet splitting (Fig.\ref{STO-t2g}) by tetragonal elongation for low-spin state $Fe^{3+}$ ion are$\cite{PilbrowESR}$:
\begin{eqnarray}
\label{BasisTetra}
 {|\psi,\pm \rangle = a|t_2(0),\pm\rangle ~\pm~ib|t_2(\pm 1), \mp \rangle }   
\end{eqnarray}
where
\begin{eqnarray}
\label{BasisTetraAB}
\lefteqn{a = {\frac{1}{\sqrt{2}}\bigg\lbrack 1 + \frac{\eta -\frac{1}{2}}{(\eta^2 - \eta +\frac{9}{4})^{1/2}}\bigg\rbrack^{1/2}}}\nonumber\\&&
{b = {\frac{1}{\sqrt{2}}\bigg\lbrack 1 - \frac{\eta -\frac{1}{2}}{(\eta^2 - \eta +\frac{9}{4})^{1/2}}\bigg\rbrack^{1/2}}}
\end{eqnarray}
Where spin up or down is indicated by the $\pm$ or $\mp$ signs. The energy difference between $|t_2(\pm 1)\rangle$ and $|t_2(0)\rangle$ are assumed as $\delta$ $(Fig.\ref{STO-t2g})$. One electron spin-orbit coupling parameter $\zeta$ has been used in the ratio $\eta = \frac{\delta}{\zeta}$.
No initial splitting is apparent with fitting line in the characteristics graph ($Fig.~\ref{STOg1}$) due to electronic quadrupole fine structure term $D=0$ with effective spin $S=\frac{1}{2}$. Also, the eigenstate $|\psi,\pm \rangle$ leads to principal $g$-factors with orbital reduction factor $k$ as$\cite{PilbrowESR}$:
\begin{eqnarray}
\label{BasisTetraAB1}
\lefteqn{g_{\scriptscriptstyle\parallel} = g_e(a^2-b^2)-2kb^2}\nonumber\\&&
{g_{\scriptscriptstyle\perp} =g_ea^2+2\sqrt{2}kab}
\end{eqnarray}
The parameters $a$ and $b$ can be given simply with anisotropy angle $\phi$ with z-axis as$\cite{AbragamESR}$~$a=sin\phi$ and $b=cos\phi$ (the $g_{\scriptscriptstyle\perp}$ value differs in sign in this treatment). 
 
\begin{figure}[t!] 
\includegraphics[width=3.5in]{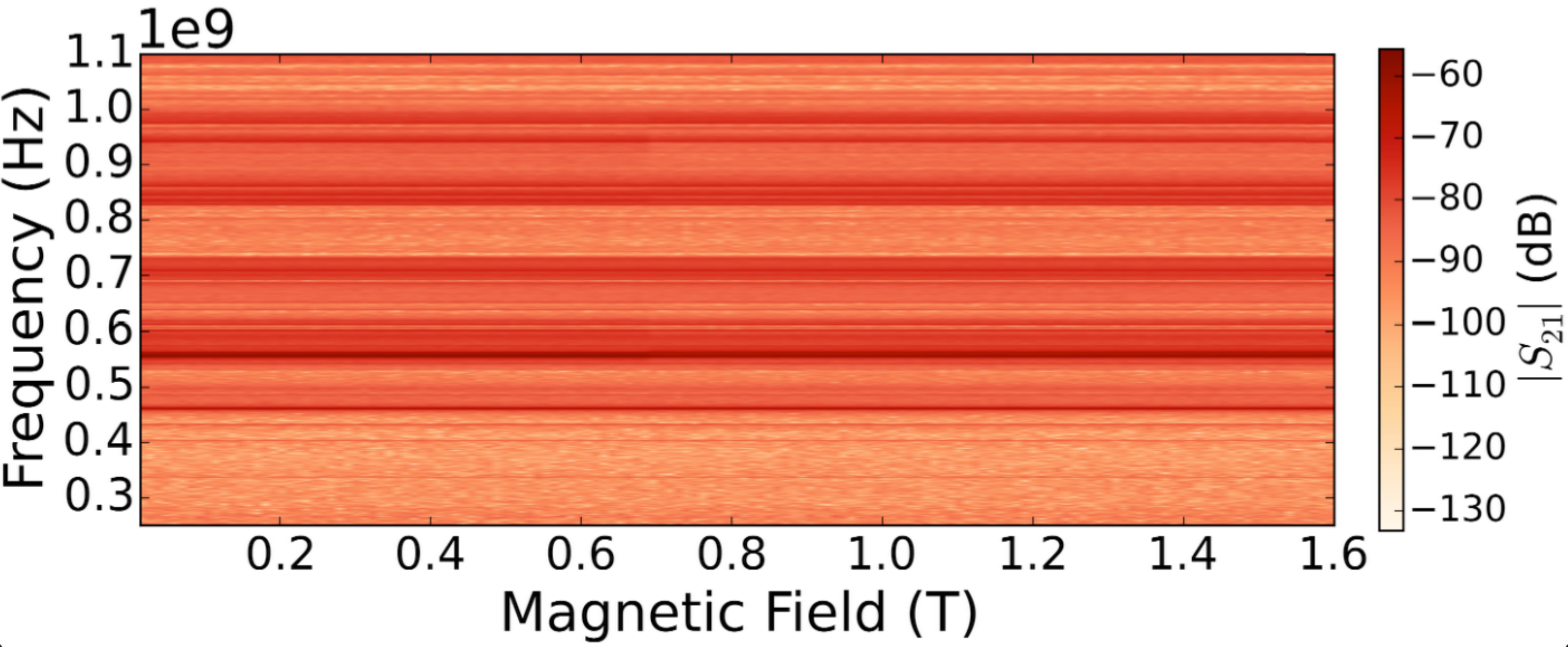}
\caption[ Density plot of the mode spectrum within the frequency range of 0.25 GHz to 1.10 GHz at high magnetic field]{\label{plot59H} Density color plot of transmission $S_{21}$ of the mode spectrum within the frequency range $0.25~GHz$ to $1.10~GHz$ in $SrTiO_3~(STO)$ crystal at $20~mK$ temperature. No $ESR$ spectrum is observed in the magnetic field 0.015 T to 1.6 T.}
\end{figure}

With the variation of the angle $\phi$ from $0$ to $\frac{\pi}{2}$ the parallel $g$-factor $g_{\scriptscriptstyle\parallel}$ becomes equal to free electron value. The values of $\eta << 0$ implies to a ground state with $g_{\scriptscriptstyle\perp Fe}\approx 0$ (equation-$\eqref{BasisTetraAB}$ and $\eqref{BasisTetraAB1}$). The parameter $k$ is defined with Fermi-contact term $k_\circ \simeq 0.43$ as $\alpha^2 k_\circ$ where $\alpha^2$ is 1 for completely ionic bond, and if the overlap integral is vanishingly small then $\alpha^2$ is $0.5$ for completely covalent bond$\citep{BondingESR}$. It gives us information about impurity ion's covalency in the site. Within this condition of bonding, there is no possible fitting value of $\eta$ according to equation-$\ref{BasisTetraAB}$ and $\ref{BasisTetraAB1}$ matching with the measured parallel g-factor $g_{\scriptscriptstyle\parallel F}=5.51$. In fact this treatment is feasible for the estimation of tetragonal distortion of the octahedron, and not for in presence of iron-oxygen-vacancy $(Fe^{3+}-V_O)$ in the octahedron where $Fe^{3+}$ ion is displaced away from the oxygen vacancy $V_O$ by about $0.35$~\r{A}$\cite{STO-FeVoEPR1}$. As a supporting results of this matter, in our ESR spectroscopy measurement of low-spin $Fe^{3+}$ trivalent ion the in $SrLaAlO_4$ crystal $FeO_6$ octahedron at $20~mK\cite{SLA}$, the considered value of $k=0.33$ is the fitted with the value $\eta\approx -1$ according to the measured g-factor $g_{\scriptscriptstyle\parallel Fe}=2.028$ in the tetragonal anisotropy$\cite{AbragamESR,DefectCu^2}$. The reason is that, in case of $Fe^{3+}$ in $SrLaAlO_4$, the trivalent $Al^{3+}$ ion is  substituted by this trivalent $Fe^{3+}$ ion without any possibility of iron-oxygen-vacancy $(Fe^{3+}-V_O)$ in the octahedron. Also, an axial electron paramagnetic resonance (EPR) spectrum in iron doped (cubic) strontium titanate has been observed at $3.3$ and $1.85$ cm wavelength with effective g-values $ g_{\scriptscriptstyle\parallel}=5.99$ and $5.96$ respectively$\cite{STO-FeEPRChargComp}$ in measurements with the constant magnetic field parallel to [001] of the crystal (these g-factor were perpendicular when the magnetic field is parallel to [100] of the crystal).  The spectrum is attributed to $Fe^{3+}$ ion in a strongly tetragonal electronic crystalline field produced by local charge compensation at a nearest neighbour oxygen site due to iron-oxygen-vacancy $(Fe^{3+}-V_O)$ in the octahedron.

High field ESR spectroscopy does not show any ESR transition line (Fig.\ref{plot59H}). One reason is due to the extremely high permittivity of $SrTiO_3$ at low temperatures, so the specimen becomes over-moded at high frequency. For example, the selected modes from $4.5~GHz$ to $11~GHz$ at room temperature were lowered to the frequency range of $0.44~GHz$ to $1.12~GHz$ at $20~mK$ as the relative permittivity increased from about $300$ to a value of the order of $10^4$. Secondly, the Q-factor of the modes were only of the order of $10^3$ which is not good-enough to detect other impurity ions of very dilute concentration. Also, at this low Q-factor, it is not  possible to observe ESR transition line of $Fe^{3+}$ isotopes which have abundance $\leq 2\%$ of the ion concentration. So, only the iron element of atomic mass $56$ is the source of $Fe^{3+}$ ion of nuclear spin $I=0$. As a result low-field ESR transitions are confined eventually to $F=S$ as a spin half transitions with selection rule $|\Delta M_F|=|\Delta M_S|=1$. 

Although the asymmetry of iron-oxygen-vacancy $(Fe^{3+}-V_O)$ is attributed to the dispersion of ESR line$\cite{STO-EsrCriAsyMuller,STO-EsrFE-Muller1}$, the observed results shows distinct (without overlapping) ESR line with little variation with the decrease of frequencies (Fig.\ref{sweep454},\ref{sweep599} and \ref{sweep622}). The trend of asymmetry is significant down to P-band frequency (Fig.\ref{STOg1}).These transition lines would show mirror reflection to the right in rotational measurement. M$\ddot{u}$ller et al analysed the spectrum in the tetragonal $SrTiO_3$ at a temperature $T\to T_c$ showing mirror reflection of the ESR line in rotational measurement$\cite{STO-EsrCriAsyMuller}$. In fact, the change of $Fe^{3+}$ ion center-symmetry in the $FeO_6$ complex as a soft-mode characteristics of ferroelectric phase transition and the influences of iron-oxygen-vacancy $(Fe^{3+}-V_O)$, interactively sensitive to the asymmetry in the octahedral rotational parameter $\Phi$ in $SrTi0_3$ and thus shows this mirror reflection of ESR line.
 
 Also in addition to the tetragonal elongation, in the millikelvin temperature experiment of $SrTiO_3$ crystal, some lower order crystal symmetry like rhombic distortion $(R)$ $(Fig.~\ref{STO-t2g})$ is apparent. So, the term $E(S_x^2 + S_y^2)$ need to be considered. In such a rhombic distortion in the asymmetry of iron-oxygen-vacancy $(Fe^{3+}-V_O)$, it is observed that the ESR line in the Fig.\ref{sweep622},\ref{sweep599} and \ref{sweep454} comparatively a little broader in cascade as the decrease of $ESR$ transition frequency. M$\ddot{u}$ller et al's analysis with experiment supports this observation. This effect implies to a decrease of effective g-factor from 5.51 to 5.32 (Fig.\ref{STOg1}).

\begin{acknowledgements}

This work was funded by Australian Research Council (ARC), Grant no. CE170100009. Thanks to Dr. Jeremy Bourhill for assistance with data acquisition and Mr. Steve Osborne for technical support. 
\end{acknowledgements}

\subsection{References:}

\end{document}